# Development of production technology of nanostructure monodisperse powder of different substances

O. Khazamov, M. Ramadanov

(Daghestan State Technical University, Russia, Makhachkala, 367014)

**Abstract**

One of the breakthrough areas of physical science over the past decade has been research in the field of nanostructured materials of various substances, almost all of the properties of which differ sharply from the properties of macroscopic size. It is difficult to list the areas of science and technology, where nanomaterials are not implemented due to their new original properties. In this regard, the development of technologies for the production of nanoscale structures is a topical issue which is of great interest in almost all areas of science and industry. As we know, all the known technologies of nanostructured materials have limited scope, and are inefficient. [1,2] The main purpose of the work is the definition of technological parameters of dispersion sustainable modes of various groups of substances.

## I. Introduction

The proposed technology for producing nanostructured materials is a more versatile and highly productive. Research in this field of science at the Department of Physics, Dagestan State Technical University started in 1988 [3,5], and the research team of faculty of the University has achieved some success.

The dispersion of the liquid is made between two electrodes. One electrode is a capillary, which is fed by a high potential, and the second electrode is disk-shaped and grounded through microammeter. Capillary system in the plane of the liquid-phase material is fed through the PTFE tube at the other end of which is fixed tank with liquid being investigated. The fluid level in the reservoir varies relative to the end of the capillary.

## II. Development of technology

Initial studies were conducted for the case when the liquid level equal to that of the capillary. In this state at the top of the capillary is observed hemisphere of the liquid. After the capillary being fed by potential, the elongated hemisphere of fluid appears on the end of capillary at a certain value of potential, when the surface tension is unable to keep the drop - the failure of the droplets occurs, and they move to the second electrode. During flight breaking drops take spherical shape. With the growth of the potential on the capillary, the separation frequency of droplets increases and further transition to the stationary dispersion state in the form of streams occurs. With further increase of potential, the flow transits in an unstable state, and it splits into two streams, which are surrounded by a misty stream in the form of a cone.

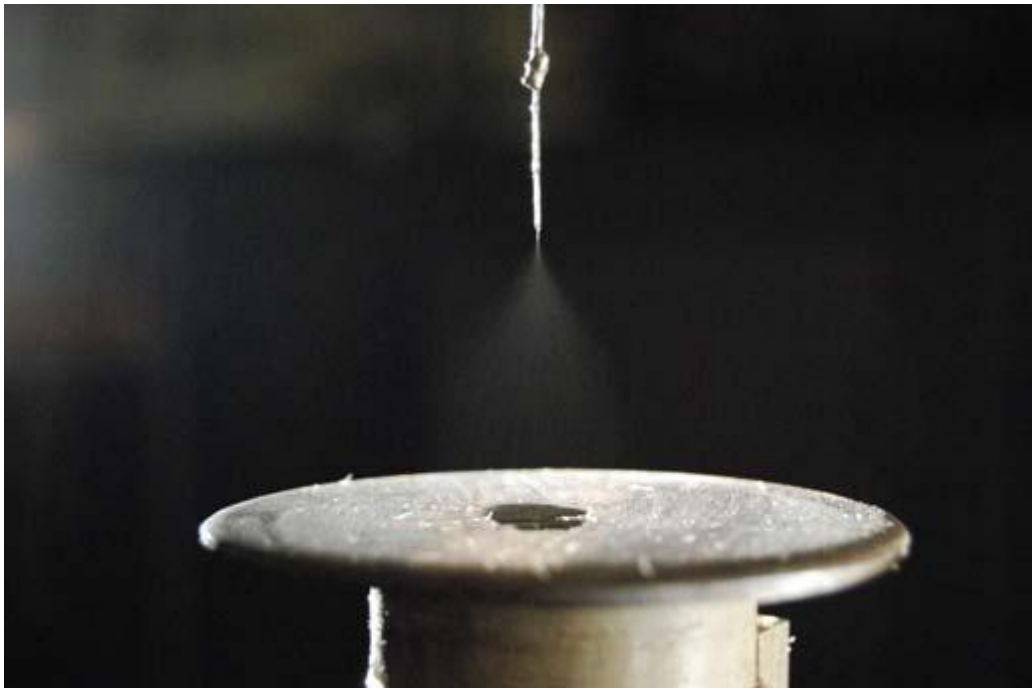

Fig.1 The photo of homogeneous steady monodisperse stream

Further growth of potential leads to a rotation of jets around the axis of the capillary. A subsequent increase in the capacity leads to the disintegration of jets on larger quantity. Next, jets merge and monostructure nanoparticles flow formed. To visualize the area of dispersion it was illuminated by the intense luminous flux. The subsequent increase in potential leads to a gradual transition of flow to polydisperse structure; in the flow of particle particles appears, sizes of which vary from several nanometers to several micrometers. For butyl alcohol were made volt-current-voltage characteristics for different modes depending on the distance between the electrodes and the liquid level above the tip of the capillary. (The results are given on the fig.2 )

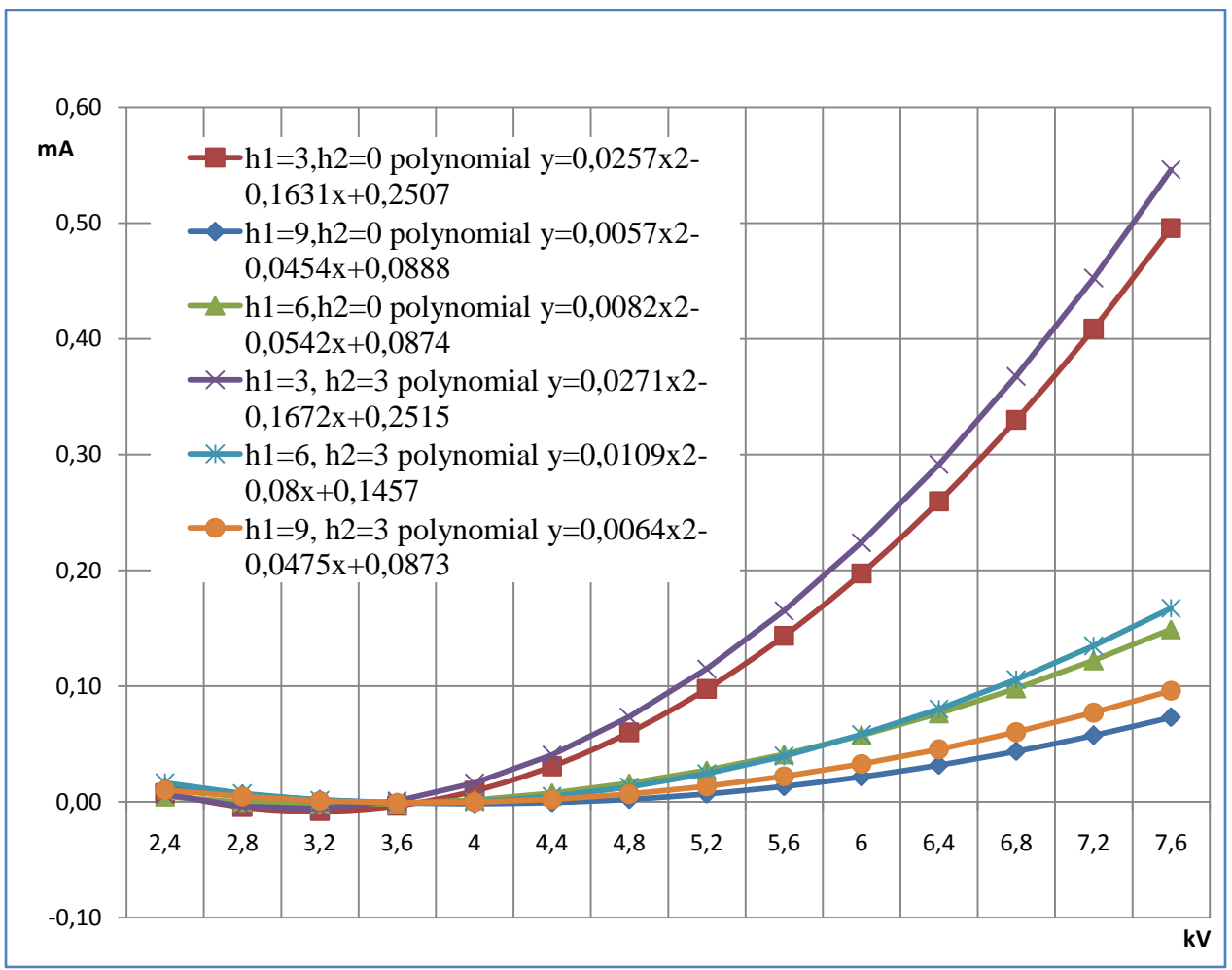

Fig.2 Current-voltage characteristics for butyl alcohol.

In the next part of comprehensive studies, the results of experimental investigations of current-voltage characteristics for transformer oil were obtained. (Fig.3)

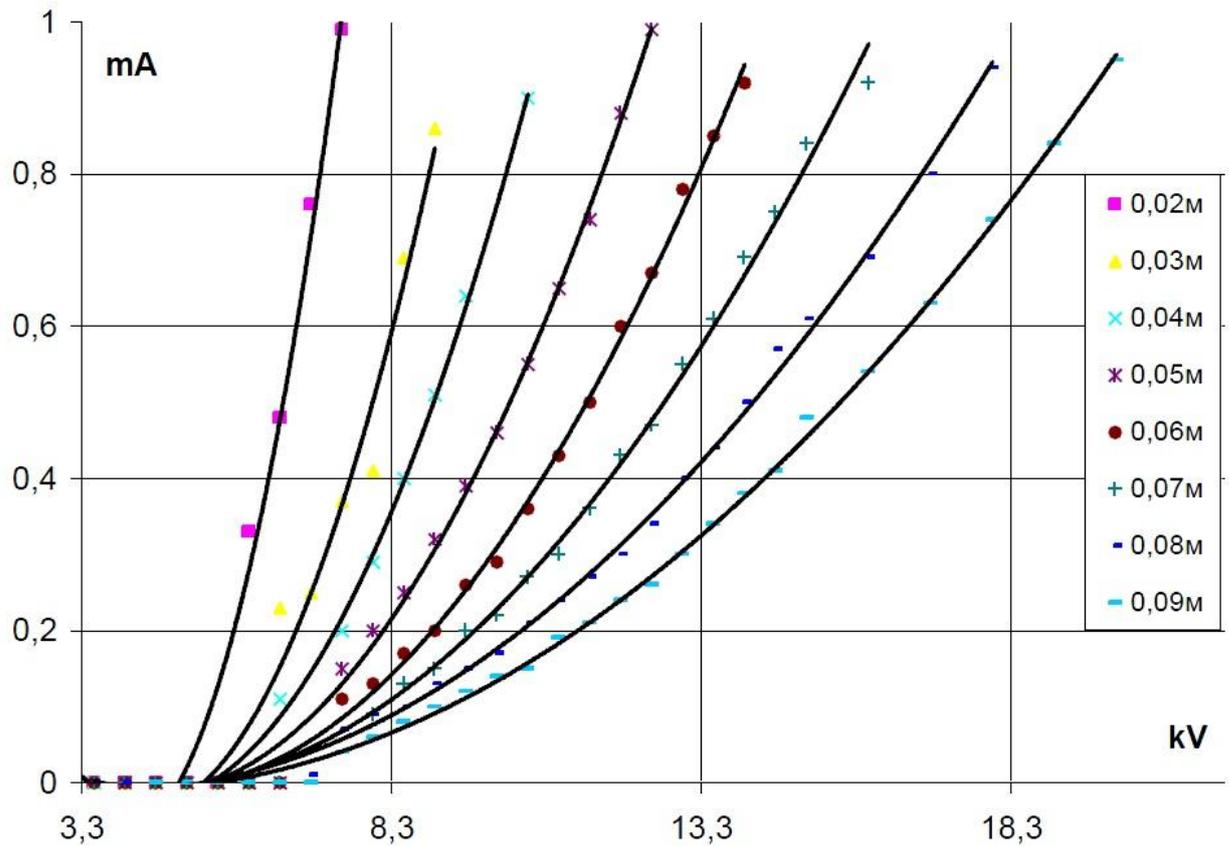

Fig.2 Current-voltage characteristics for for transformer oil.

As a result of experimental studies particle sizes of saturated hydrocarbons were identified with NTEGRA PRIMA complex for scanning probe microscopy with the processing program NOVA, and it was the established that the size of produced particles vary in the range of $2*10^{-9} \div 12*10^{-9}$ m.

The Photo of wax powder on coverslips is shown in Figure 3.

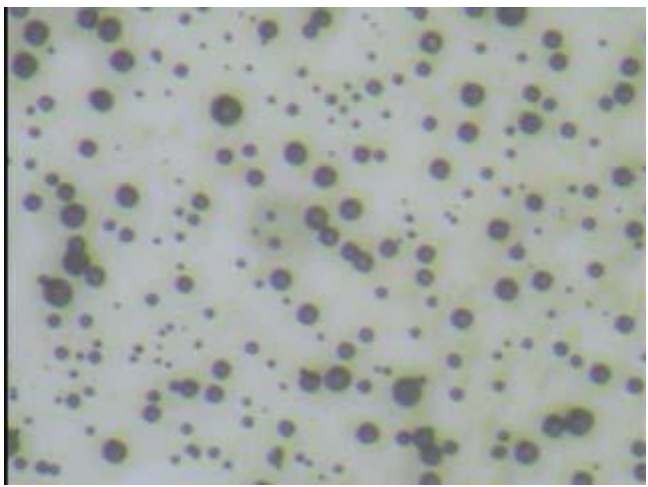

Fig. 3. The photo of wax powder on coverslips

The diameter of fine particles measured by microscope is $2 * 10^{-9}$ m.

## III. Conclusions

The experimental results confirm the scope of the proposed method for nano-materials, in particular all saturated hydrocarbons. To further narrow range of particle sizes dispersed flow is divided by the size of particles in the various fields of force.

During the pilot studies modes of dispersion of the investigated substances with monodisperse nanoscale structure flows were fixed. Using the proposed technology powders of various substances with the required size of the particles can be obtained. It was established that the regimes of stable monodisperse flows can be obtained for almost all investigated substances.

The studies were conducted with the financial support of the "START"; the State Contract number 4979 r/7232 from 30.03.2007.